# Transport and entry of plasma clouds/jets across transverse magnetic discontinuities:

## three-dimensional electromagnetic particle-in-cell simulations

Gabriel Voitcu[1] and Marius Echim[1,2]

[1]*Institute of Space Science, Magurele, Romania*

[2]*Belgian Institute for Space Aeronomy, Brussels, Belgium*

**Corresponding author:** G. Voitcu, Institute of Space Science, Atomistilor 409, P.O. Box: MG-23, Magurele 077125, Ilfov, Romania (gabi@spacescience.ro)

**Key points**

Perpendicular transport of plasma clouds/jets is investigated through PIC simulations

Penetration takes place for an initial momentum large enough to overcome the magnetic barrier

The cloud/jet is braked and eventually stopped in the non-uniform magnetic fields

**Abstract**

In this paper we use three-dimensional electromagnetic particle-in-cell simulations to investigate the interaction of a small-Larmor radius plasma cloud/jet with a transverse non-uniform magnetic field typical to a tangential discontinuity in a parallel geometry. The simulation setup corresponds to an idealized, yet relevant, magnetospheric configuration likely to be observed at the magnetopause during northward orientation of the interplanetary magnetic field. The numerical simulations are adapted to study the kinetic effects and their role on the transport and entry of localized plasma jets similar to those identified inside the Earth's magnetosheath propagating towards the magnetopause. The simulations reveal the formation of a polarization electric field inside the main bulk of the plasma cloud that enables its forward transport and entry across the transverse magnetic field. The jet is able to penetrate the transition region when the height of the magnetic barrier does not exceed a certain critical threshold. Otherwise, the forward transport along the injection direction is stopped before full penetration of the magnetopause. Moreover, the jet is pushed back and simultaneously deflected in the perpendicular plane to the magnetic field. Our simulations evidence physical processes advocated previously by the theoretical model of impulsive penetration and revealed in laboratory experiments.

## 1. Introduction

The interaction of solar wind with the terrestrial magnetosphere is a key topic in space physics and space weather science. Studying the transport of plasma across the magnetopause for a broad range of interplanetary conditions and by considering the full physics is fundamentally important to better understand the interaction between the solar wind and the terrestrial magnetosphere. The mechanisms that favour access of the solar wind plasma into the magnetosphere and the role played by both the geomagnetic field and the interplanetary magnetic field are not completely elucidated.

In the last decades the increased resolution of scientific instruments on-board magnetospheric missions enabled systematic detection of localized magnetosheath plasma irregularities like, for instance, those reported by *Hietala et al.* (2012), *Karlsson et al.* (2012) or *Savin et al.* (2012). Some of these magnetosheath structures (called also plasma elements, clouds, jets, streams, blobs, plasmoids, density or dynamic pressure enhancements) were observed deep inside the terrestrial magnetosphere by various spacecraft around the Earth, e.g. *Lundin and Aparicio* (1982), *Woch and Lundin* (1991, 1992), *Yamauchi et al.* (1993), *Lundin et al.* (2003), *Lu et al.* (2004), *Dmitriev and Suvorova* (2012), *Gunell et al.* (2012), *Shi et al.* (2013).

More recently, *Archer and Horbury* (2013) and *Plaschke et al.* (2013) performed statistical analyses of THEMIS data to characterize the dynamical properties of a large number of high-speed jets identified inside the magnetosheath. It has been shown that these structures are characterized by strong density and velocity enhancements with respect to the background plasma. Most of the observed jets are propagating anti-sunward in the frontside magnetosheath and are likely to interact with the magnetopause. *Gunell et al.* (2014) used *in-situ* measurements performed by Cluster spacecraft in the magnetosheath to study the interaction of high-speed plasmoids with the magnetopause and their penetration into the magnetosphere. A statistical analysis of penetrating jets presented recently by *Dmitriev and Suvorova* (2015) shows that a large fraction (>60%) of the jets detected by THEMIS in the magnetosheath do penetrate the magnetopause and enter the magnetosphere. These authors also suggest that the penetrating jets have in general velocities larger than a given threshold (~200 km/s) and exhibit electrodynamic properties consistent with the impulsive penetration (IP) model (*Lemaire*, 1977).

Several physical mechanisms have been proposed to explain the transfer of mass, momentum and energy at the magnetopause, such as: magnetic reconnection, Kelvin-Helmholtz instability, kinetic Alfvén waves or impulsive penetration. In a recent review of plasma transport and entry, *Wing et al.* (2014) show that all aforementioned mechanisms provide similar entry rates of solar wind plasma into the Earth's plasma sheet under northward interplanetary magnetic field (IMF) conditions. Nevertheless, each of these mechanisms has its limitations and further investigations are required to clarify this issue.

The transport of plasma across magnetic fields is by its own nature a three-dimensional electromagnetic process. Indeed, there is (i) convection along the injection direction, *Ox*, (ii) electric self-polarization along the perpendicular direction to the magnetic field and the forward plasma bulk velocity, *Oy*, and (iii) expansion along the magnetic field direction, *Oz*. While streaming across the transverse magnetic field, different classes of electromagnetic waves are generated even for low-beta plasma structures, as discussed, for instance, by *Neubert et al.* (1992). These authors used electromagnetic kinetic simulations to study the motion of low-beta plasma clouds, emphasizing the presence of electromagnetic waves with different frequencies during the transport across the background magnetic field. Even though the electromagnetic approach provides a more complete solution on this problem, the dynamics of low-beta plasmas across transverse magnetic fields can be investigated also with electrostatic simulations, as summarized in Table 2.

In this paper we investigate the kinetic effects and their role on the propagation of localized plasma elements across non-uniform transverse magnetic fields by using three-dimensional electromagnetic particle-in-cell (PIC) simulations. We analyze the interaction of small Larmor radius plasma clouds with background magnetic field profiles typical to tangential discontinuities (TDs). The simulation scenario considered here corresponds to a magnetosheath plasma irregularity (or blob, cloud, jet, plasmoid) streaming towards the terrestrial magnetopause and interacting with it, in a simplified magnetospheric configuration corresponding to a northward IMF orientation.

The paper is organized as follows. In the second section we describe the simulation setup considered. In the third section we analyze the simulation results obtained and discuss the kinetic effects related to the transport of three-dimensional plasma clouds across transverse magnetic barriers. The fourth section includes our summary and discussion, while the last one emphasize the final conclusions of the paper.

## 2. Simulation setup

The numerical simulations are performed in a three-dimensional geometry that allows the simultaneous investigation of the plasma cloud electrodynamics in all relevant directions. The interaction with the background plasma is not included in this study as we wish to investigate the interaction of the cloud with the background magnetic field. A schematic diagram of the simulation setup is shown in Figure 1.

The steady-state background magnetic field is parallel to the *z*-axis and perpendicular to the initial bulk velocity of the plasma element. Its magnitude increases linearly between two asymptotic states over a finite-width transition region that covers few ion Larmor radii:

$$B_0(x) = \begin{cases} B_1, \text{ for } x < x_1 \\ B_1 + \dfrac{\delta B}{\delta x}(x - x_1), \text{ for } x_1 \le x \le x_2 \\ B_2, \text{ for } x > x_2 \end{cases} \quad (1)$$

where: $\delta x = x_2 - x_1$ is the width of the transition region and $\delta B = B_2 - B_1$ is the magnetic field increase from its asymptotic value at the left side of the transition region, $B_1$. This type of magnetic profile has been found in kinetic models of tangential discontinuities (e.g. *Sestero*, 1964, 1966; *Lemaire and Burlaga*, 1976; *Roth et al.*, 1996) and it is a steady-state Vlasov-Maxwell equilibrium solution at the interface of two plasma populations with different densities and temperatures. The moving plasma cloud/jet simulated in our study can be seen as a perturbation to the equilibrium solution. We assume that the background steady-state magnetic field profile is established prior to the injection of the plasma cloud into the simulation domain and that this background configuration remains unchanged during the entire simulation runtime. Similar profiles of the magnetic field have been previously used in numerical simulations with test-particles to study the dynamics of charged particle systems in prescribed configurations of the electromagnetic field (i.e. *Echim and Lemaire,* 2003; *Echim,* 2004; *Voitcu and Echim,* 2012; *Voitcu et al.*, 2012). The self-consistent magnetic field due to the internal plasma currents computed during the simulation runtime adds to the background magnetic field (1). Initially, the electric field is set to zero in the entire simulation domain.

The electrons and protons forming the three-dimensional plasma cloud are uniformly distributed over a rectangular region localized at the left hand side of the transition area, as shown in Figure 1. Their positions are overlapping such that the net charge density inside the simulation domain is equal to zero at the beginning of the simulation. The initial velocity distribution function (VDF) of both plasma species is a displaced Maxwellian with the average velocity, $\vec{U}_0$, parallel to the positive $x$-axis and perpendicular to the background magnetic field given by equation (1). The velocities of electrons and protons are initialized according to their corresponding displaced Maxwellian distribution function. Thus, at $t = 0$, the plasma cloud is injected into the simulation domain with an initial bulk velocity pointing towards the discontinuity surface; the injection takes place at the left hand side of the transition region where the magnetic field is uniform. We performed simulations for the case of small Larmor radius plasma clouds, i.e. plasma structures characterized by a transversal dimension much larger than the ion Larmor radius.

The numerical simulations shown in the present paper are performed using a modified version of the TRISTAN code (*Buneman*, 1993) adapted to the interaction of localized plasma structures with transverse magnetic fields. In the present study, the boundary conditions are assumed to be periodic for both particles and fields. To minimize the possible undesired influence of boundaries on the plasma dynamics, we try to keep the edges of the simulation domain as far as possible from the localized plasma element. A detailed description of the PIC-3D code and the boundary conditions used in our simulations can be found in *Voitcu* (2014); for details about the original TRISTAN code see *Buneman* (1993) and *Cai et al.* (2003).

The input parameters for the simulations shown in the present paper are given in Table 1. We focus on three different configurations/cases: (i) case A, further called *"open magnetic barrier"*, corresponding to a thin, gradual varying field of a penetrable TD, (ii) case B, further called *"wide magnetic barrier"*, corresponding to a thick and impenetrable TD and (iii) case C, further called *"closed magnetic barrier"*, corresponding to a thin, steep and impenetrable TD. The width of the TD is of the order of few ion Larmor radii, in agreement with theoretical arguments (e.g. *Willis*, 1978; *Roth et al.*, 1996) and observations (e.g. Cluster measurements, see for instance *Haaland et al.*, 2004) of the magnetopause thickness. For all three cases considered in this study the plasma cloud is characterized by a high dielectric constant and its transversal dimension is considerably larger than the ion Larmor radius. The total simulation time covers 1.5 ion Larmor periods or equivalently 54 electron Larmor periods. It should be mentioned that such localized plasma structures with high dielectric constant and small Larmor radius are frequently observed inside the terrestrial magnetosheath (e.g. *Karlsson et al.*, 2012; *Plaschke et al.*, 2013). Most of them are streaming towards the magnetopause and are likely to interact with it. A recent study indicates that 60% of the jets detected in the equatorial regions of the magnetosheath by THEMIS between 2007 and 2008 do penetrate the magnetopause and enter into the magnetosphere (*Dmitriev and Suvorova*, 2015). The main goal of our work is to investigate the physical mechanisms that enable the transport and entry of plasma jets, similar to the ones observed experimentally, that interact with a simplified, yet relevant, magnetopause-like configuration.

Table 2 compares the input parameters of particle-in-cell simulations performed to study the plasma motion across transverse magnetic fields and published in the literature in the last two decades: the two-dimensional electrostatic simulations of *Livesey and Pritchett* (1989) and *Galvez and Borovsky* (1991), the three-dimensional electromagnetic simulations of *Neubert et al.* (1992) and the three-dimensional electrostatic simulations by *Gunell et al.*

(2009). The last two lines of the table correspond to the simulations described in this paper and respectively to the experimental observations of the magnetosheath high-speed jets by Cluster and THEMIS spacecraft (*Karlsson et al.*, 2012; *Plaschke et al.*, 2013). The numerical simulations of *Livesey and Pritchett* (1989), *Galvez and Borovsky* (1991) and *Neubert et al.* (1992) have been performed for a uniform magnetic field, while the geometry used in the study of *Gunell et al.* (2009) is closely related to laboratory experiments with large Larmor radius plasma clouds entering a curved magnetic field (see also *Hurtig et al.*, 2003). To our knowledge, the present study shows the first three-dimensional particle-in-cell simulations of a small Larmor radius plasma cloud transported across a non-uniform magnetic profile typical to a tangential discontinuity (as the magnetopause).

## 3. Numerical results

The numerical simulations are performed in a three-dimensional setup illustrated schematically in Figure 1. Three different configurations of the "magnetopause" profile are tested. The transition profiles of the non-uniform background magnetic field are shown in Figure 2. In all three cases the magnetic field is parallel, only its intensity is increasing across a finite region. In this study we are interested to investigate the role of the magnetic field increase, therefore we disregard the effects of magnetic shears. The latter are subject of a subsequent study.

We consider a rectangular plasma cloud localized initially at the left hand side of the transition region and characterized by a non-zero bulk velocity in the positive direction of the *x*-axis. The initial electric field is everywhere equal to zero. The input parameters of the simulations are specified in Table 1.

All physical quantities are normalized as follows: the number density and bulk velocity are normalized to their corresponding initial values $n_0$ and $U_0$, the electric and magnetic fields are normalized to $E_0=U_0 \cdot B_1$ and $B_1$, respectively, while the time and spatial coordinates are normalized to the initial ion Larmor period $T_{Li}$ and grid spacing $\Delta x$, respectively.

### 3.1. Case A: Open magnetic barrier

The first simulation configuration describes a localized transition region where the background magnetic field increases by 10% over a finite-width area (further called discontinuity) that covers 2.3 ion Larmor radii (see the green line in Figure 2) and is uniform in the rest of the simulation domain. We consider a plasma cloud/jet whose transversal width, $w_y$, is considerably larger compared to $\delta x$, the thickness of the magnetic transition region: $w_y/\delta x \approx 8$.

The time evolution of the plasma number density is shown in Figure 3 for two different cross-sections inside the simulation domain: left panels correspond to the plane $xOy$ perpendicular to the background magnetic field, for $z$=203, and respectively right panels correspond to the plane $xOz$, for $y$=103. At $t$=0, the electrons and ions have the same number density and the net electric charge inside the plasma cloud is equal to zero. The initial shape of the plasma element is rectangular with uniform density and is localized at the left hand side of the transition region where the magnetic field is uniform. At $t$=0.75$T_{Li}$, the front side region of the cloud moved across the magnetic discontinuity. Some internal density variability is observed, a possible effect of internal waves and instabilities, but the cloud preserves its overall shape and convects as a coherent structure; for a discussion on high-frequency oscillations related to the transport of plasma clouds across magnetic barriers, see the papers of *Hurtig et al.* (2005) and *Brenning et al.* (2005). The $xOz$ cross-section shows that the cloud also expands in the direction of the background magnetic field and its density decreases accordingly (see the right column of Figure 3). At $t$=1.50$T_{Li}$, the entire cloud moves across the discontinuity and penetrates into its right hand side ("magnetospheric" side). Due to the field-aligned expansion, the density decreases further, but the transversal dimension of the cloud remains virtually identical to the initial one. The results clearly illustrate the propagation of the plasma cloud across the region of sharp magnetic field variation ("magnetopause" in the simulation setup shown in Figure 1). We call this case "open magnetic barrier". During a total travel time of 1.5 ion Larmor periods, the plasma element moved along the positive direction of the $x$-axis over a distance of about 20 ion Larmor radii, which corresponds to an average bulk velocity of approximately 90% of the initial velocity, a possible signature of braking effects.

Being more energetic, the electrons move along the $z$-axis faster than the ions and, consequently, an outward ambipolar parallel electric field is established at the boundaries of the plasma cloud, as can be seen in Figure 4. This differential expansion plays an important role in the plasma dynamics along the $z$-axis. The parallel electric field at the boundaries of the cloud is acting differently on the two plasma species and will tend to reduce its expansion rate. Indeed, the outer directed $E_z$ decelerates the electrons and accelerates the ions; the charge separation between the two species along $Oz$ decreases and the parallel electric field is significantly diminished. By comparing the red and green peaks in Figure 4, we can note that the parallel electric field at $t=0.09T_{Li}$ is approximately 2 times weaker than at $t=0.04T_{Li}$ and therefore is less efficient to inhibit the plasma expansion. Later on, a certain equilibrium is achieved and the two plasma species are expanding at equal rates. As a result, the parallel electric field vanishes at the edges of the plasma cloud. After 0.21 ion Larmor periods from the beginning of the simulation, the $E_z$ component of the electric field is approximately equal to zero everywhere along the $z$-axis (see the blue line in Figure 4). However, the plasma expansion along the magnetic field direction continues, but at a lower rate than initially. By the end of the simulation, the suprathermal particles reached the boundaries of the simulation domain. Nevertheless, the bulk of thermal electrons and ions are still located inside the center of the simulation box. The fluctuations observed for the parallel electric field, $E_z$, are due to the statistical noise introduced by the limited number of particles loaded into the simulation domain. In the absence of the outward parallel electric field, the thermal electrons would have reached the boundaries of the simulation domain after only 0.4 ion Larmor periods from initialization. The rapid evanescence of the ambipolar electric field is perhaps also related to the limited parallel expansion and the mutual interaction of waves propagating along the background magnetic field. For a detailed discussion on plasma expansion in vacuum see the work of *Galvez and Borovsky* (1991), but also the papers of *Farrell et al.* (1998) and *Birch and Chapman* (2001).

Another direct and important consequence of the parallel expansion is the significant decrease of the density that might have also an effect on the transversal motion of the plasma cloud along the $x$-axis. As shown by Figure 3, the density in the main core of the cloud is a few times smaller at $t$=1.50$T_{Li}$ than initially, $n/n_0\approx0.14$. Nevertheless, the dielectric constant is large, $\varepsilon\approx60$, and the forward propagation of the cloud is not altered by its rarefaction.

The plasma element is braked during its forward motion along the $x$-axis. To illustrate more accurately the deceleration process, we calculated the plasma bulk velocity, at the end of the simulation, by using the average velocities of both electrons, $\vec{U}_e$, and ions, $\vec{U}_i$:

$$\vec{U} = \frac{m_e n_e \vec{U}_e + m_i n_i \vec{U}_i}{m_e n_e + m_i n_i} \qquad (2)$$

where $n_e$ and $n_i$ are the electron and ion number densities. In order to avoid any unrealistically large bulk velocities that could arise in those spatial bins populated with less particles, $\vec{U}$ is calculated only for those grid cells with a density of at least 5% of $n_0$.

Figure 5 illustrates $U_x$, the forward component of the plasma bulk velocity in the same two planes presented in Figure 3, at $t$=1.50$T_{Li}$. The plasma cloud is braked while it moves downstream the magnetic discontinuity. To better illustrate this effect, we plot in the bottom panel of Figure 5 the variation profile of $U_x$ along the $y$-axis, for $x$=184 and $z$=203 (i.e. along the vertical white line in the top-left panel). The mean forward velocity at the right hand side of the transition region (marked with red) is 30% smaller than the one at the left hand side: $U_x/U_0$=0.7.

In Figure 6 we show the perpendicular components of the electric field, $E_x$ (top panels) and $E_y$ (bottom panels), in the $z$=203 (left column) and $y$=103 (right column) planes, at $t$=0.04$T_{Li}$. The approximate boundaries of the plasma element are traced in those locations where the electron and ion number densities decrease significantly such that $n/n_0\approx0.05$. A detailed view is given in Figure 7. The left panel shows the perpendicular electric field within the boundaries of the plasma cloud, while the right panel illustrates the profiles of $E_x$ and $E_y$ along the y-axis. The two components of the electric field have been averaged from $z=185$ to $z=223$ (in the left plot) and furthermore from $x=110$ to $x=150$ (in the right plot) to diminish the local fluctuations related to the particle-in-cell noise.

A perpendicular electric field pointing predominantly along the positive y-axis is established inside the main bulk of the plasma cloud shortly after injection, as illustrated by the lower panels of Figure 6 and the left panel of Figure 7. The intensity of this perpendicular electric field is not uniform, but shows significant fluctuations due to the inherent simulation noise. Nevertheless, its average value taken between $y=83$ and $y=124$ closely match the normalization factor $E_0$: $E_y/E_0=0.92$ and $E_x/E_0=-0.02$, as emphasized in the right panel of Figure 7. Thus, the only non-vanishing component of the total electric field at the left hand side of the transition region, inside the main bulk of the plasma cloud, is $E_y \approx U_0 \cdot B_1$. Note that the initial value of the electric field is set to zero in the entire simulation box.

The self-consistent electric field established inside the main bulk of the plasma cloud in the early stages of the simulation is the result of plasma polarization in the perpendicular plane to the magnetic field, as described by *Schmidt* (1960). Indeed, since the electrons and ions are gyrating in opposite directions, two space charge layers of different polarities are formed at the lateral edges of the plasma element along the perpendicular direction to both the plasma injection velocity, $\vec{U}_0$, and the background magnetic field, $\vec{B}_0$. These charge layers sustain a polarization electric field, $\vec{E}_p$, inside the quasineutral bulk of the plasma cloud/jet (*Schmidt*, 1960):

$$\vec{E}_p = -\vec{U}_0 \times \vec{B}_0 \qquad (3)$$

Since the kinetic energy of the plasma element is significantly larger than the electric field energy or, equivalently, $\varepsilon \gg 1$, the polarization electric field (3) enables further convection of the plasma element across the magnetic field with approximately the initial bulk velocity.

The perpendicular electric field is not strictly confined inside the main bulk of the plasma element, but it extends also to its nearby regions. Indeed, the $E_x$ component takes large values close to $x \approx 100$ and $x \approx 160$, the front side and the trailing edge of the cloud/jet. Higher values of the $E_y$ component are observed around $y \approx 75$ and $y \approx 130$. The detailed description of the kinetic and electromagnetic features near the edges of the plasma element are discussed in *Voitcu* (2014) and are consistent with *Galvez et al.* (1988). Far away from the cloud's boundaries, the perpendicular electric field is virtually zero.

In Figure 8 we illustrate the evolution of the $B_z$ component of the total magnetic field. Inside the boundaries of the plasma cloud/jet $B_z$ is smaller by 1% than the background magnetic field, $B_0$. Thus, a small diamagnetic cavity is formed in the current position of the plasma cloud. The $B_x$ and $B_y$ components of the magnetic field (not shown here) are significantly smaller (<0.02%) than the background field pointing along the positive $z$-axis. As expected, the self-consistent contribution of the plasma particles to the total magnetic field is negligible, since we consider a low-beta plasma cloud ($\beta$=0.12).

The impulsive penetration mechanism proposed by *Lemaire* (1977) discusses the interaction of localized magnetosheath plasma structures with planetary magnetopauses. According to this model, the convection velocity of a low-beta plasma cloud in the non-uniform magnetic field, $B_0(x)$, satisfies the following equation derived from conservation of the first adiabatic invariant (*Lemaire*, 1985):

$$\frac{m_e + m_i}{2} U_x^2(x) + (\mu_e + \mu_i) B_0(x) = \text{const.} \tag{4}$$

where $\mu_e$ and $\mu_i$ are the magnetic moments of the thermal electrons and ions. Equation (4) is valid for a large dielectric constant and shows that the cloud is slowed down while it moves across stronger magnetic fields; in other words is adiabatically braked and its forward bulk motion energy is converted into gyration energy. The critical magnetic field, $B_c$, for which all the initial convection energy is transformed into gyration energy and the plasma element is virtually stopped is given by (*Lemaire*, 1985):

$$B_c = \frac{(m_e + m_i)U_0^2 + m_e V_{Te0}^2 + m_i V_{Ti0}^2}{2(\mu_e + \mu_i)} \tag{5}$$

where $V_{Te0}$ and $V_{Ti0}$ are the initial thermal velocities of the electrons and ions. When the plasma element reaches the critical point $x_c$ where $B_0(x_c)=B_c$, the entire convection energy is transformed into gyration energy and the forward motion stops. For the input parameters used in our simulations the critical magnetic field (5) is $B_c=1.23B_1$.

The process of adiabatic braking described by *Lemaire* (1985) for space plasmas has been verified in the past by various laboratory studies. Indeed, numerous experiments performed over the years revealed the formation of Schmidt's polarization electric field and the adiabatic braking of localized plasma elements interacting with transverse magnetic fields (e.g. *Bostick*, 1956; *Wetstone et al.*, 1960; *Demidenko et al.* 1967, 1969, 1972; *Wessel et al.*, 1988). *Echim and Lemaire* (2005) showed that the two-dimensional boundaries of such localized plasma structures (plasmoids, jets, clouds) are characterized by a non-vanishing parallel electric field, due to the parallel gradient of the perpendicular velocity and/or kinetic pressure, that decouples the plasma element from the background and sustain its transport across the transverse magnetic field. For an extensive review of the impulsive penetration mechanism, see *Echim and Lemaire* (2000).

In our numerical experiments (i) the plasma-beta parameter is low ($\beta=0.12$), (ii) the plasma dielectric constant is large ($\varepsilon=500$) and (iii) the magnetic field variation over an ion Larmor radius is small ($\sim 0.04B_1$). The simulations show the adiabatic braking of the cloud/jet while it advances into the region of stronger magnetic field. According to equation (4) of the IP model, the plasma bulk velocity at the right hand side of the magnetic discontinuity would be $U_x=0.74U_0$, in good agreement with the results of our PIC simulations where $U_x=0.70U_0$.

### 3.2. Case B: Wide magnetic barrier

In the second simulated case the transition region is approximately 7 times wider than in the previous one, while the magnetic field gradient remains unchanged. More exactly, the background magnetic field increases by 67% over 16 ion Larmor radii (see the blue line in Figure 2). This time the transversal width of the plasma element is comparable with the thickness of the transition region, i.e. $w_y/\delta x \approx 1.2$.

In Figure 9 we show the time evolution of the number density in the plane normal (left column) and parallel (right column) to the background magnetic field, at three different moments of time: $t=0.50T_{Li}$ (top panels), $t=1.00T_{Li}$ (middle panels) and $t=1.50T_{Li}$ (bottom panels). Figure 10 illustrates the $U_x$ component of the plasma bulk velocity in the direction normal to the discontinuity for exactly the same planes and times. In order to avoid large numerical errors, the bulk velocity is computed from equation (2) for those grid cells that have a number density at least 5% of the initial value. After 1.5 ion Larmor periods from the beginning of the simulation, the plasma element moved almost completely inside the region of non-uniform magnetic field, as shown in Figure 9. Its forward convection velocity has decreased considerably while advancing into the stronger magnetic field (as illustrated by the top and bottom panels of Figure 10).

 The braking of the plasma cloud is illustrated more clearly in Figure 11 where we show the variation of $U_x$ along the y-axis at $t=0.50T_{Li}$ (blue curve), $t=1.00T_{Li}$ (green curve) and $t=1.50T_{Li}$ (red curve), for the x coordinates indicated by black lines in Figure 10. At $t=0.50T_{Li}$, the forward convection velocity in the front side region of the plasma cloud, at $x=168$, is 29% smaller than the initial velocity: $U_x/U_0=0.71$. Later on, at $t=1.50T_{Li}$, the plasma bulk velocity decreases to $U_x=0.16U_0$ and the forward motion is significantly braked. The critical value of the magnetic field for which the plasmoid would stop according to (5) is reached in $x_c=173$. Nevertheless, as shown in Figure 9, the plasma element is able to move beyond this point. By the end of the simulation, the plasma cloud is considerably slowed down and almost completely stopped before leaving the transition region. These results confirm that the adiabatic braking operates for plasma elements injected in non-uniform magnetic fields.

During the forward propagation into the stronger magnetic field, the plasma element is compressed along the convection direction. At the end of the simulation, its length along the $x$-axis is ~33% smaller than its initial $x$-size (as indicated by the top-left panel in Figure 3 and the bottom-left panel in Figure 9). This compression effect is related to the braking of the plasma element. Indeed, while streaming across the transition region, the forward convection velocity inside the plasma cloud is varying along the $x$-axis (see the bottom panels of Figure 10). Thus, the bulk velocity of the leading edge is smaller than that of the trailing edge. The rear regions are streaming faster than the frontal ones and the plasma element is compressed. It should be mentioned that this effect is also observed in case A where is less efficient since the magnetic field increase across the transition region is smaller (as shown by the top and bottom panels of Figure 3), leading to a length of the plasma element at $t$=1.50$T_{Li}$ only ~25% smaller than at $t$=0. Thus, broader transition regions compress more the impacting plasmoids/clouds/jets.

### 3.3. Case C: Closed magnetic barrier

In this simulation the gradient of the magnetic field is 5 times stronger than in the previous two simulations. More precisely, the background magnetic field increases by 50% over the same distance as in the first case (see Figure 2).

The dynamics of the plasma element is illustrated in Figures 12 and 13 and show the electron number density and the forward component of the bulk velocity in the two different cross-sections defined for the previous cases and for simulation instances $t$=0.75$T_{Li}$ (top panels of Figure 12) and $t$=1.50$T_{Li}$. In this case, although the plasmoid is injected with the same initial bulk velocity as in the previous cases, it cannot fully enter the transition region. Indeed, the forward motion along the $x$-axis is fully stopped – the magnetic barrier is "closed" in this case. Indeed, from $t$=0.75$T_{Li}$ to $t$=1.50$T_{Li}$, the front-side edge of the cloud, localized at $x$≈170, is virtually at rest and the main bulk of the plasma is not able to cross the magnetic discontinuity and move further into its right hand side (compare the top and bottom panels of Figure 12). Moreover, it is strongly deflected along both positive and negative directions of the $y$-axis. A closer look into the top panels of Figure 13 reveals the backward propagation of the plasma cloud at the end of the simulation ($U_x$<0 in the vicinity of the TD). This feature is emphasized more clearly in the bottom panel of Figure 13, which shows that the plasma element is not only stopped, but also repelled and pushed back along the negative $x$-axis, away from the magnetopause. The lateral "wings" of the plasma element, localized around $y$≈75 and $y$≈160, exhibit a quite large negative bulk velocity: $U_x$≈−0.35$U_0$.

The motion of the electrons and ions inside the discontinuity is determined by the electric and gradient-B drifts acting on the two plasma species. For the typical geometry used in our simulations, the only non-vanishing component of the gradient-B drift velocity is oriented along the $y$-axis:

$$V_{\nabla B}^{y} = \frac{W_{\perp}}{qB_0^2} \cdot \frac{\delta B}{\delta x} \qquad (6)$$

where $W_{\perp}$ is the gyration energy in the perpendicular plane to the magnetic field and $q$ is the electrical charge. The gradient-B drift deflects the electrons in the $-Oy$ direction, while the ions drift in the opposite direction. As a result, the structure of the two space charge layers formed at the lateral edges of the plasma cloud in the early stages of the simulation, prior to the interaction the magnetic discontinuity, is affected by the differential effect of the gradient-B drift in the vicinity of the cloud's boundaries. Ions are continuously accumulating at the "top" negative layer localized at higher $y$-values, while electrons, at the "bottom" positive layer localized at lower $y$-values. The net effect is that the space charge layers are gradually neutralized by the particles moving with the gradient-B drift velocity. Note that, since the gradient-B drift velocity is proportional to the gyration energy, only the most energetic electrons and ions are effectively contributing to the "net neutralization" of the two layers. In time, as the plasma cloud is advancing into the magnetic discontinuity, the total net electrical charge inside both space charge layers is diminished and the corresponding polarization electric field, $E_y$, gets weaker. If the magnetic field increase is sufficiently large, the polarity of the two boundary layers reverses and the polarization electric field turns from positive to negative values. Consequently, the forward convection velocity, given by the $x$-component of the electric drift velocity ($U_{E,x}=E_y/B_z$), also switches sign and turns negative. Thus, the plasma element is stopped and pushed backwards, away from the magnetopause, along the $-x$-axis. In this case, the magnetopause-like magnetic discontinuity/barrier remains closed and cannot be penetrated by the incoming plasma element/cloud/jet.

The gradient-B drift plays also an important role for the global dynamics of the plasma element along the $y$-axis. Indeed, the differential effect of the gradient-B drift on the two charged species sustains the formation of a polarization electric field in the $x$-direction, normal to the discontinuity, in the frontal regions of the plasma cloud. This electric field component is responsible for the symmetric deflection of the particles along both positive and negative directions of the $y$-axis, as shown by the bottom-left panel of Figure 12. Note that this effect is also observed in cases A and B where is much less strong. A detailed analysis of the plasma deflection along the $y$-axis and the possible consequences on the plasma transport at the magnetopause is treated in a subsequent paper under preparation.

In the left panel of Figure 14 we show the $E_y$ component of the electric field in the $z$=203 cross-section inside the simulation domain, at $t$=1.50$T_{Li}$, but only for those grid cells having a number density of at least 5% from its initial value. The corresponding zero-order (electric) drift velocity, $U_{E,x}=E_y/B_z$, is given in the right panel of Figure 14. Inside and nearby the magnetic discontinuity $E_y \leq 0$ and $U_{E,x} \leq 0$. Also, the lateral "wings" of the plasma element, localized around $y \approx 65$ and $y \approx 160$, are characterized by a strong negative electric field and electric drift velocity. The electric field fluctuations observed within the current position of plasma element are most probably related to simulation noise; see *Hurtig et al.* (2005) and *Brenning et al.* (2005) for discussions on high-frequency oscillations related to the transport of plasma across magnetic barriers.

When one compares the right panel of Figure 14 with the top-left panel of Figure 13, one can see that the values of the normal component of the electric drift (or MHD) velocity, $U_{E,x}$, are in general larger than those of $U_x$, derived from the first order moment of the velocity distribution function. This effect is more evident at the lateral "wings" of the plasma element emphasized by the two black dashed ovals in the right panel of Figure 14. Indeed, the average value of $U_{E,x}$ at the "bottom" edge of the plasma cloud ($y \approx 65$) is 32% larger than $U_x$. At the opposite edge ($y \approx 160$), the difference is smaller, i.e. $U_{E,x}/U_x = 1.06$. Note that similar findings have been obtained by kinetic modelling (*Echim and Lemaire*, 2005; *Echim et al.*, 2005) and reported from spacecraft observations (*Lundin et al.*, 2005). These previous works show that sharp boundaries of the order of the Larmor radius are sites where the ideal MHD convection velocity, given by the electric drift, is decoupled from the actual plasma bulk velocity derived from the real moments of the velocity distribution function.

## 4. Summary and discussion

We report results obtained by three-dimensional electromagnetic particle-in-cell simulations performed to investigate the propagation of a small-Larmor radius plasma element/cloud/jet across a region of non-uniform magnetic field typical to a tangential discontinuity. We have defined a simulation setup that enables the simultaneous examination of the plasma electrodynamics along the injection direction (*x*-axis), the self-polarization along the normal direction to the magnetic field and the convection velocity (*y*-axis) and parallel expansion along the magnetic field lines (*z*-axis).

The dynamics of localized plasma elements streaming across uniform transverse magnetic fields has been studied also in the past by various numerical experiments using electrostatic one-dimensional (e.g. *Galvez*, 1987; *Cai and Buneman*, 1992) and two-dimensional (e.g. *Galvez et al.*, 1988; *Livesey and Pritchett*, 1989; *Galvez and Borovsky*, 1991; *Cai and Buneman*, 1992) particle-in-cell codes, but also electromagnetic three-dimensional PIC simulations (e.g. *Neubert et al.*, 1992). The computer experiments performed emphasized the self-polarization of plasma elements and their propagation across the transverse magnetic field, as described by *Schmidt*'s model (*Schmidt*, 1960). Later on, *Hurtig et al.* (2003) and *Gunell et al.* (2009) have studied the motion of three-dimensional plasmoids across curved magnetic fields using electrostatic particle-in-cell simulations. It should be mentioned that these numerical experiments were limited by the computing resources available at the epoch and none of them succeeded to evaluate the interaction of a plasma blob with a tangential discontinuity.

Nevertheless, the interaction of a plasma irregularity with a tangential discontinuity was investigated with ideal, Hall and resistive MHD, as well as with hybrid simulations (see the review by *Echim and Lemaire*, 2000). Although the MHD simulations are not fully adapted to simulate the kinetic effects that play a key role for plasma transport across magnetic fields, they pointed out that two-dimensional infinitely long diamagnetic plasma filaments move across a tangential discontinuity when their magnetization is either parallel or antiparallel to the asymptotic magnetic field inside the magnetosphere (*Dai and Woodward*, 1994, 1995, 1998; *Huba*, 1996). Resistive MHD simulations suggest that the jet can "partially" penetrate for reduced magnetic shear across the discontinuity (*Ma et al.*, 1991). Hybrid simulations (*Savoini and Scholer*, 1994) show the importance of ion kinetics in the phenomenological description of impulsively penetrating jets and evidence asymmetric propagation of the jet inside the magnetosphere. In addition to the truncated geometry, another major limitation of this class of numerical simulations is that they do not quantify the effect of space charge layers and the role of the self-polarization.

The background magnetic field used in our simulations is increasing over a finite-width transition with the thickness of few ion Larmor radii. The plasma cloud is injected into the simulation domain at the left hand side of the discontinuity or transition region, with an initial bulk velocity pointing normal to the discontinuity surface. We discussed three cases: (i) thin and penetrable barrier, (ii) thick and impenetrable barrier and (iii) thin and impenetrable barrier. For all three cases the plasma dielectric constant is large ($\varepsilon$=500) and the beta parameter is small ($\beta$=0.12). The simulation setup used here illustrates an idealized, yet relevant, magnetospheric geometry during northward IMF, designed to study the kinetic effects and their role on the transport of localized plasma clouds/jets, similar to those identified inside the Earth's magnetosheath, across a transverse magnetic barrier, as the magnetopause.

The numerical results obtained revealed the formation of a polarization electric field inside the main bulk of the plasma element. Indeed, due to the opposite gyration directions of the electrons and ions, two space charge layers are built on either sides of the plasma cloud along the *y*-axis. The two boundary layers with different polarities sustain the polarization electric field oriented perpendicular to the magnetic field and plasma convection velocity. This electric field is established self-consistently in the very early stages of the simulation ($t<0.1T_{Li}$). Its intensity and orientation match the theoretical description of *Schmidt* (1960), i.e. $E_y=U_x \cdot B_z$. The polarization electric field plays an important role for the plasma dynamics prior to the interaction with the discontinuity since it enables the forward convection of the plasma cloud across the transverse magnetic field. If the plasma dielectric constant is large enough (here $\varepsilon$=500), the convection motion along the *x*-axis is maintained at the same velocity as the initial one.

The interaction of the plasma element with the discontinuity is controlled by the height of the magnetic barrier, i.e. the magnetic field at the right hand side of the transition region. The simulations show that the plasma element achieves full entry across the discontinuity if the magnetic field at the right hand ("magnetospheric") side of the transition region is less than a critical threshold determined by the plasma temperature and dynamical pressure, as shown by equation (5). In this case the magnetic barrier is "open" and the plasma element is able to move across it. As illustrated in cases A and B, the convection motion into the larger magnetic field is efficiently slowed down by the gradual conversion of the bulk motion energy into gyration energy, consistent with the adiabatic braking advocated by laboratory and theoretical studies (e.g. *Demidenko et al.*, 1972; *Lemaire*, 1985). If the magnetic field at the right hand side of the discontinuity is too large, the cross transport along the *x*-axis is stopped and the plasma element cannot penetrate the transition region; the magnetic barrier is "closed". Moreover, the cloud is pushed back and simultaneously deflected along the positive and negative directions of the *y*-axis, as illustrated in case C.

The braking process emphasized in our simulations is related to the gradient-B drift acting within the magnetic discontinuity. The differential deflection of the electrons and ions along the *y*-axis tends to neutralize the two space charge layers. Thus, the corresponding polarization electric field weakens and the forward motion of the plasma element is slowed down. When the magnetic field is large enough the polarity of the two layers reverses, leading to the reversal of both the polarization electric field and the plasma convection velocity along *x*-axis.

Simultaneously with the cross propagation along the *x*-axis, the plasma element is expanding rapidly in the directions parallel and antiparallel to the background magnetic field. The expansion diminishes the density of the plasma element; at the end of the simulations, the main core of the plasma element is ~7 times more tenuous than initially. The numerical experiments performed here revealed the formation of a parallel electric field close to the edges of the plasma element along the *z*-axis. This electric field reduces the very fast thermal expansion of the particles at the early stages of the simulation. In time, the expansion rates of the two plasma species tend to equalize and the electric field vanishes. Note that in the absence of such a parallel electric field, the thermal electrons would have reached the boundaries of the simulation domain after only 0.4 ion Larmor periods from initialization.

Since in this paper we consider a low-beta plasma cloud, our numerical results could be partially reproduced with electrostatic particle-in-cell simulations. Indeed, the self-consistent plasma contribution to the total magnetic field is negligible and has little influence on the cloud's dynamics across the magnetic discontinuity in the cases discussed above. The cloud's transverse motion is mainly driven by electrostatic effects. Nevertheless, the electromagnetic approach gives a quantitative evaluation of the amplitude of the magnetic field perturbation produced by the plasma cloud's internal currents and confirms a-posteriori its negligible effect. However, the magnetic perturbation increases for highly diamagnetic clouds (large beta) and in this case the electromagnetic approach is recommended.

Our simulations are consistent with key features of the impulsive penetration mechanism proposed to explain the transport of solar wind plasmoids across the magnetopause. Note that, to our knowledge, this is for the time when three-dimensional particle-in-cell simulations are used to investigate the transport of small Larmor radius plasma clouds across transverse non-uniform magnetic fields, in a simplified, yet relevant, magnetopause-like configuration typical for northward IMF. We evidenced physical processes advocated previously by theoretical models and revealed in laboratory experiments. The simulations demonstrate that the entry and transport of magnetosheath jets/clouds during northward IMF is mainly regulated by the dynamic and kinetic pressure of the incoming plasma, its polarizability (dielectric constant) and the height of the magnetic barrier at the interface with the magnetosphere − the magnetopause.

The numerical simulations discussed in this paper correspond to the parallel configuration treated by the MHD simulations, thus favorable for penetration. The results obtained here reveal additional phenomena not accessible to MHD approaches. Indeed, our magnetopause has a finite thickness and is not infinitesimal as in MHD simulations. Together with the self-consistent electric field generated by self-polarization, we show that the forward motion of the cloud is braked in an increasing magnetic field due to the continuous transfer of energy from bulk motion into gyration. Our simulations reveal that contrary to the ideal MHD predictions, the cloud is irreversibly stopped if the magnetospheric field is strong enough. An additional set of simulations that consider shears of magnetic field at the magnetopause has been performed and is the topic of a future publication. The inclusion of background plasma is also envisaged in our future simulations.

## 5. Final conclusions

In the present paper we investigate the transport of magnetosheath-like plasma irregularities across transverse magnetic discontinuities in a simplified magnetospheric configuration likely to be observed at the magnetopause during northward IMF orientation.

For this purpose we use three-dimensional electromagnetic particle-in-cell simulations. Here are the main findings of our study:

- A self-polarization electric field is established inside the main bulk of the plasma element in the very early stages of the simulation. This self-consistent electric field plays a key-role in our simulations since it enables the forward propagation of the plasma cloud across the transverse magnetic field.

- The plasma element is able to penetrate the transition region, i.e. the magnetopause in our simulation setup, when the height of the magnetic barrier does not exceed a certain critical threshold for a given dynamical pressure of the plasma element – the magnetic barrier is "open". In this case the plasma element achieves full entry inside the right hand side of the transition region, i.e. the magnetosphere. While streaming into the larger magnetic field, the convection motion is significantly slowed down, consistent with the impulsive penetration mechanism. The plasma cloud is completely stopped when the magnetic field at the right hand side of the discontinuity is too large – the magnetic barrier is "closed". In this case the penetration of the transition region is not possible, the plasma element being pushed back and also deflected in the perpendicular plane to the magnetic field.

- Seen the other way around, for a fixed magnetic jump at the magnetopause, only the plasmoids with larger dynamical pressure (including a larger component of the component of the bulk velocity normal to the magnetopause) will be able to penetrate.

- The plasma is expanding rapidly along the background magnetic field direction and an ambipolar parallel electric field is formed close to the edges of the cloud. This electric field reduces the very fast thermal expansion of the electrons and ions at the early stages of the simulation. Later on, the expansion rates of the two species tend to equalize and the parallel electric field vanishes.

**Acknowledgements**

The authors acknowledge support from the European Community's Seventh Framework Programme through grant agreement no. 313038/2012 (STORM) and also from the Romanian Ministry of Education and Research through projects 229EU/2013 (STORM) and 83/2013 (TIMESS). Marius Echim acknowledges the support of the Belgian PAI network CHARM and of ISSI Bern team led by Simon Wing and Jay Johnson "Plasma transport and entry into the plasma sheet".

The simulation data used to produce all the plots included in this paper can be requested by sending an e-mail to Gabriel Voitcu at one of the following addresses: gabi@spacescience.ro or gabriel.voitcu@gmail.com.

**Table 1.** Input parameters for all three simulation cases discussed here: $m_i/m_e$ is the ion-to-electron mass ratio; $U_0$ is the initial plasma bulk velocity; $V_{Ti}=(2k_BT_i/m_i)^{1/2}$ is the ion thermal speed; $\beta$ is the plasma-beta parameter (including dynamic and thermal plasma pressure); $\varepsilon$ is the plasma dielectric constant; $\tau$ is the total simulation time; $T_{Li}$ is the ion Larmor period; $\Delta x$ is the grid spacing; $\lambda_D$ is the electron Debye length; $\Delta t$ is the time-step; $c$ is the speed of light in vacuum; $N_c$ is the number of particles per grid-cell; $\delta B$ is the increase of the magnetic field across the TD of width $\delta x$; $B_1$ is the asymptotic magnetic field at the left hand side of the TD; $r_{Li}$ is the ion Larmor radius for thermal particles; $w_x$, $w_y$, $w_z$ are the widths of the plasma element along $Ox$, $Oy$, $Oz$; $mx$, $my$, $mz$ are the number of grid-cells along $Ox$, $Oy$, $Oz$.

| | $m_i/m_e$ | $U_0/V_{Ti}$ | $\beta$ | $\varepsilon$ | $\tau/T_{Li}$ | $\Delta x/\lambda_D$ | $c\Delta t/\Delta x$ | $N_c$ |
|---|---|---|---|---|---|---|---|---|
| **All cases** | 36 | 2.4 | 0.12 | 500 | 1.5 | 2.5 | 0.5 | 200 |

| | $\delta B/B_1$ | $\delta x/r_{Li}$ | $w_x/r_{Li}$ | $w_y/r_{Li}$ | $w_z/r_{Li}$ | $m_x$ | $m_y$ | $m_z$ |
|---|---|---|---|---|---|---|---|---|
| **Case A** | 10% | 2.3 | | | | | | |
| **Case B** | 67% | 16 | 19 | 19 | 12 | 255 | 205 | 405 |
| **Case C** | 50% | 2.3 | | | | | | |

**Table 2.** Selection of particle-in-cell simulations investigating the transport of localized plasma elements across transverse magnetic fields: L89 = *Livesey and Pritchett* (1989), G91 = *Galvez and Borovsky* (1991), N92 = *Neubert et al.* (1992) and G09 = *Gunell et al.* (2009); the last two lines correspond to the present paper (PP) and to observational data inside the terrestrial magnetosheath (OBS). The simulation parameters shown are: $m_i/m_e$ − ion-to-electron mass ratio; $U_0$ − initial plasma bulk velocity; $V_{Ti}$ − ion thermal speed; $\beta$ − plasma-beta parameter; $\varepsilon$ − plasma dielectric constant; $w_\perp$ − transversal width of the plasma element; $r_{Li0}$ − ion Larmor radius for $U_0$; $\tau$ − total simulation time; $T_{Li}$ − ion Larmor period; the last two columns indicate the magnetic field model used (U = uniform, C = curved, N = non-uniform) and the corresponding PIC code (ES = electrostatic, EM = electromagnetic).

|  | $m_i/m_e$ | $U_0/V_{Ti}$ | $\beta$ | $\varepsilon$ | $w_\perp/r_{Li0}$ | $\tau/T_{Li}$ | *B-model* | *PIC code* |
|---|---|---|---|---|---|---|---|---|
| **L89** | 100 | 10 | $\ll 1$ | 1600 | 0.05 | 0.35 | U | 2D-ES |
| **G91** | 100 | 11 | $\ll 1$ | 33 | 5 | 1.50 | U | 2D-ES |
| **N92** | 16 | 2.3 | $2 \cdot 10^{-2}$ | 7 | 6 | 2.00 | U | 3D-EM |
| **G09** | 92 | – | $8 \cdot 10^{-4}$ | 757 | $\leq 1.1$ | 0.70 | C | 3D-ES |
| **PP** | 36 | 2.4 | 0.12 | 500 | 8 | 1.50 | N | 3D-EM |
| **OBS** | 1836 | ~1–5 | ~1–10 | ~$10^6$–$10^7$ | ~10–50 | – | – | – |

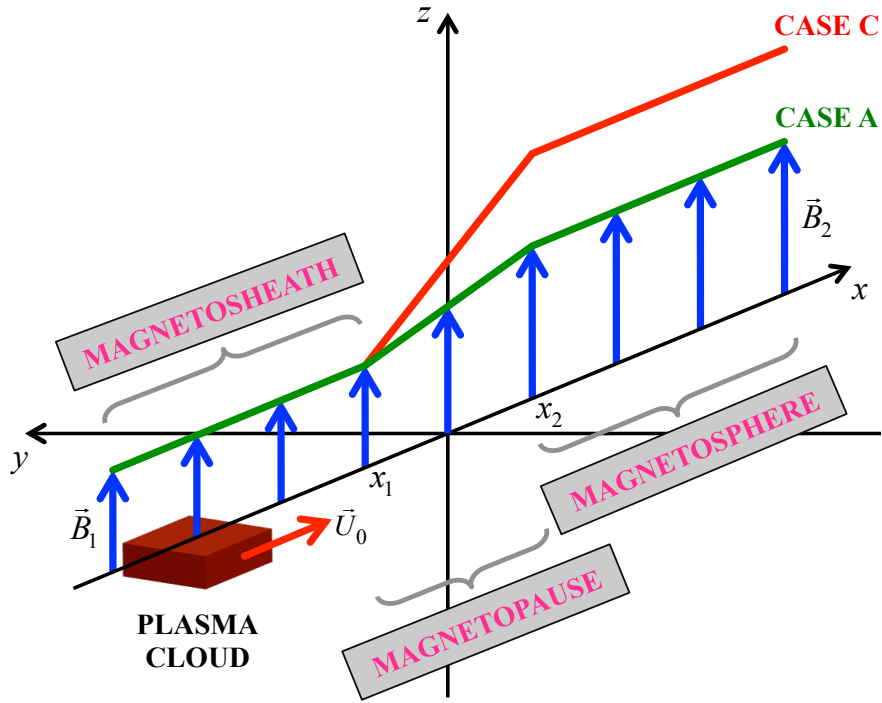

**Figure 1.** Schematic diagram of the simulation setup. The three-dimensional plasma cloud (brown rectangular box) is injected with a non-zero bulk velocity (red arrow) across a non-uniform background magnetic field (blue arrows) typical to a tangential discontinuity. This simulation geometry corresponds to a simplified, yet relevant, magnetospheric configuration for studying the interaction of localized magnetosheath plasma clouds/jets with the terrestrial magnetopause. No background plasma is taken into account in the present work.

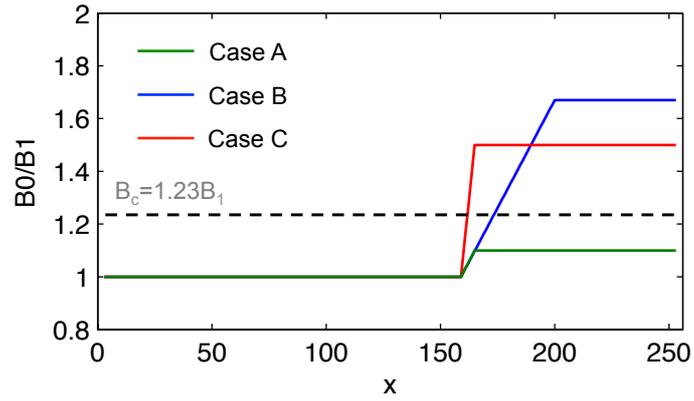

**Figure 2.** Background magnetic field profiles for case A (green line), case B (blue line) and case C (red line). The width of the transition region covers $2.3r_{Li}$ in cases A−C and $16r_{Li}$ in case B; $r_{Li}$ is the ion Larmor radius at the left hand side of the transition region. Note that the background magnetic field is oriented along $+Oz$ everywhere inside the simulation box. The horizontal dashed line marks the critical magnetic field given by equation (5) for which the forward motion of the plasma cloud along the positive $x$-axis is stopped.

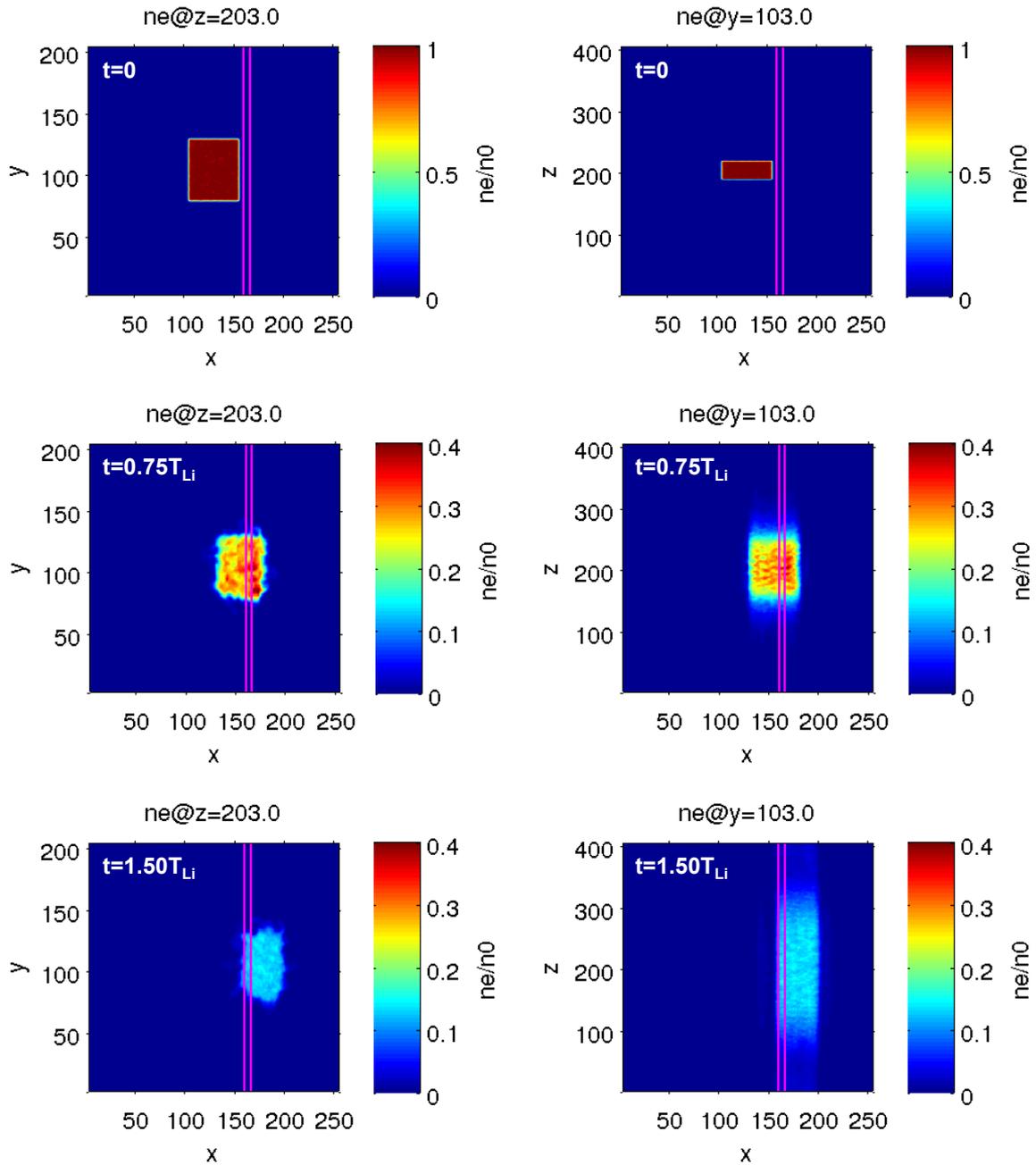

**Figure 3.** Case A: Number density of electrons in the $z$=203 (left column) and $y$=103 (right column) planes, at $t$=0 (top panels), $t$=0.75$T_{Li}$ (middle panels) and $t$=1.50$T_{Li}$ (bottom panels); $T_{Li}$ is the ion Larmor period at the left hand side of the transition region. The two magenta lines mark the boundaries of the transition region.

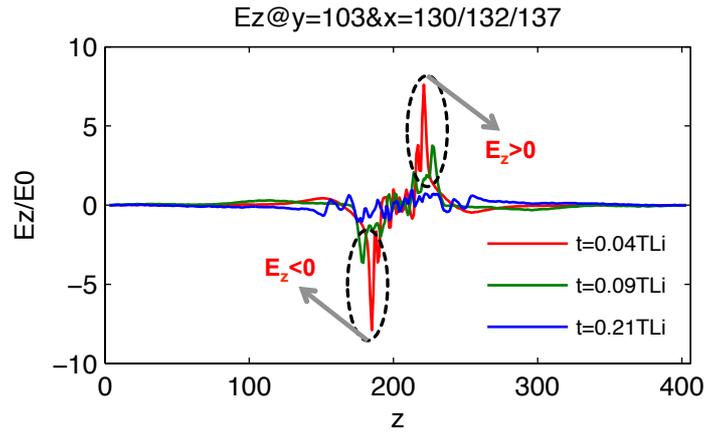

**Figure 4.** Case A: Variation profile of the parallel electric field, $E_z$, along $Oz$, at $t$=0.04$T_{Li}$ (red line), $t$=0.09$T_{Li}$ (green line) and $t$=0.21$T_{Li}$ (blue line), for $y$=103 and $x$=130 (red line), $x$=132 (green line) and $x$=137 (blue line); $T_{Li}$ is the ion Larmor period at the left hand side of the transition region. The two dashed ovals illustrate the regions close to the boundaries of the plasma element along the $z$-axis.

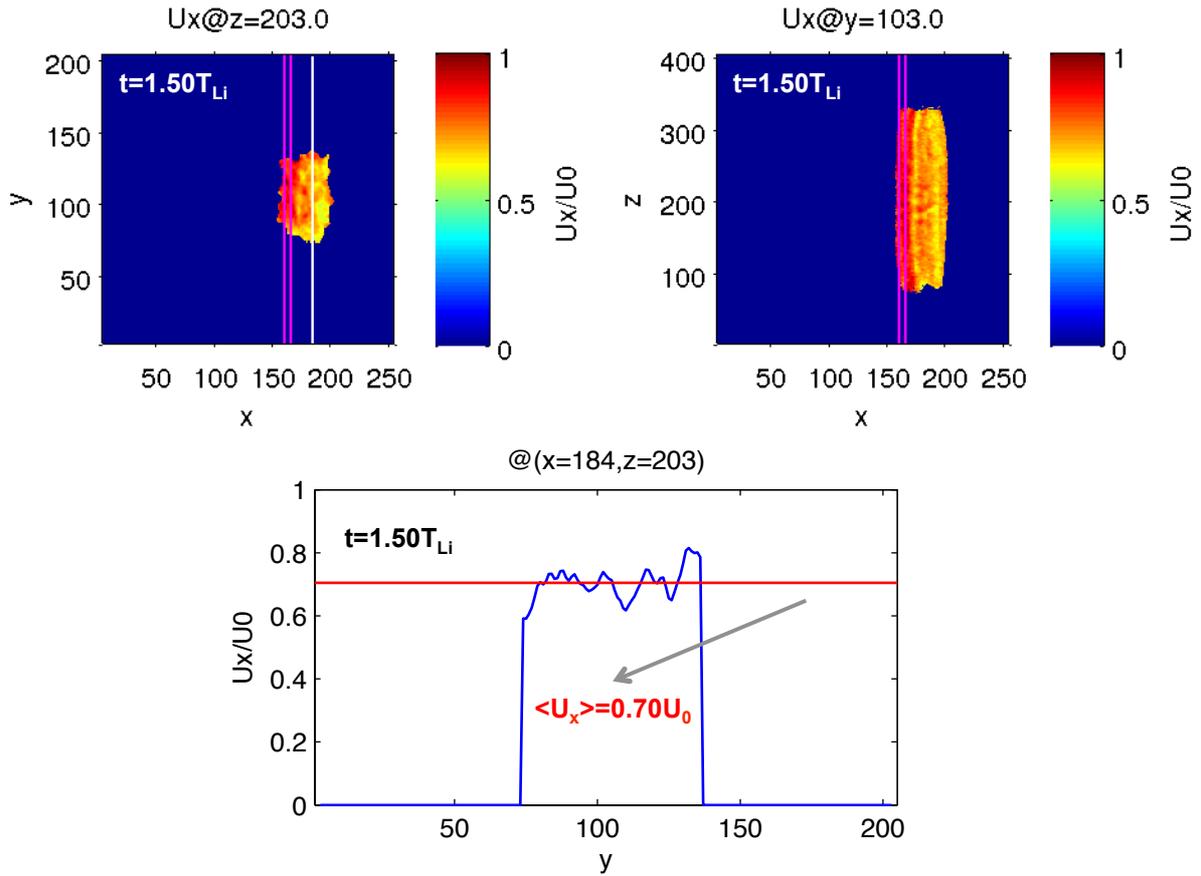

**Figure 5.** Case A: Forward component, $U_x$, of the plasma bulk velocity in the $z$=203 (top-left panel) and $y$=103 (top-right panel) cross-sections inside the simulation domain, at $t$=1.50$T_{Li}$; $T_{Li}$ is the ion Larmor period at the left hand side of the transition region. The two magenta lines mark the boundaries of the transition region, while the white line is drawn in $x$=184. The bottom panel shows the variation profile of $U_x$ along the $y$-axis, for $x$=184 and $z$=203, at $t$=1.50$T_{Li}$. The red line indicates the mean value of the variation profile shown in blue.

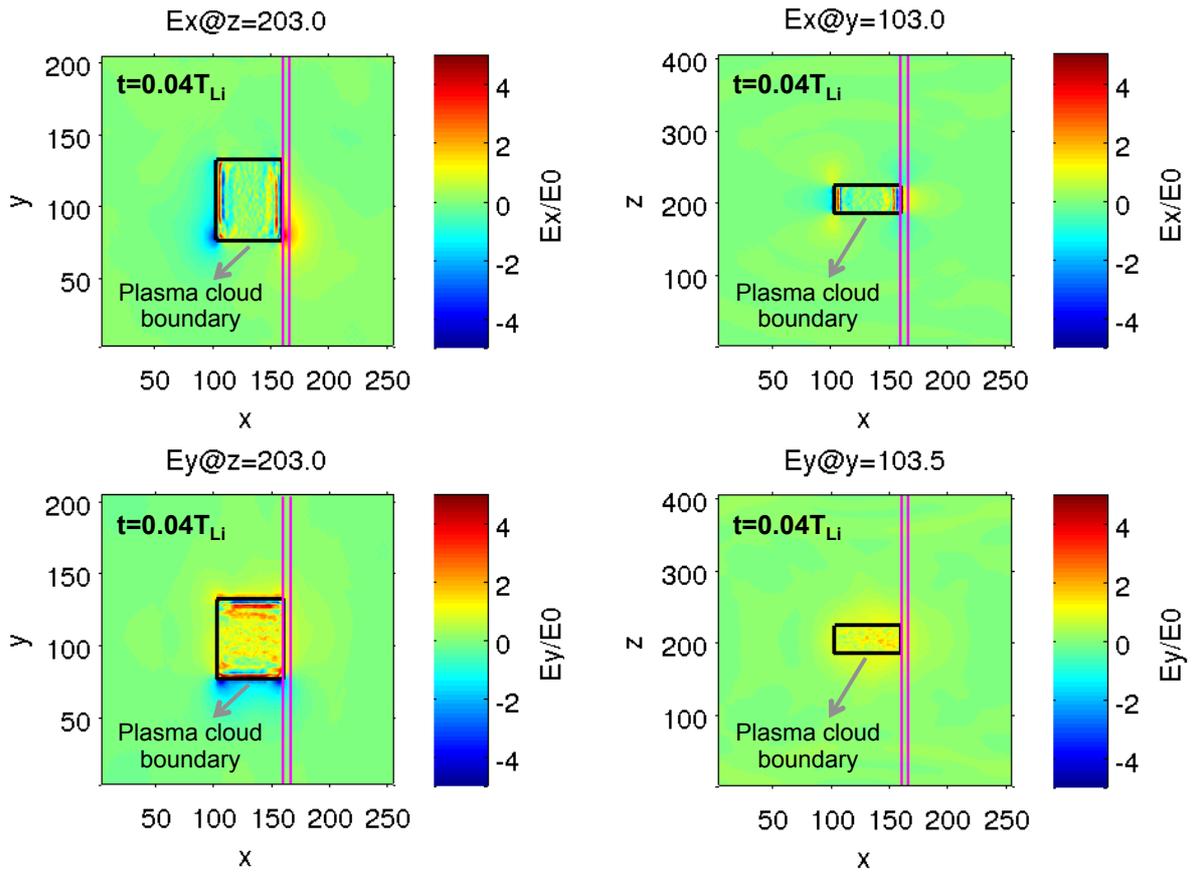

**Figure 6.** Case A: Perpendicular components of the electric field, $E_x$ (top panels) and $E_y$ (bottom panels), in the $z$=203 (left column) and $y$=103 (right column) planes, at $t$=0.04$T_{Li}$; $T_{Li}$ is the ion Larmor period at the left hand side of the transition region. The two magenta lines mark the boundaries of the transition region, while the black rectangles illustrate the boundaries of the plasma element.

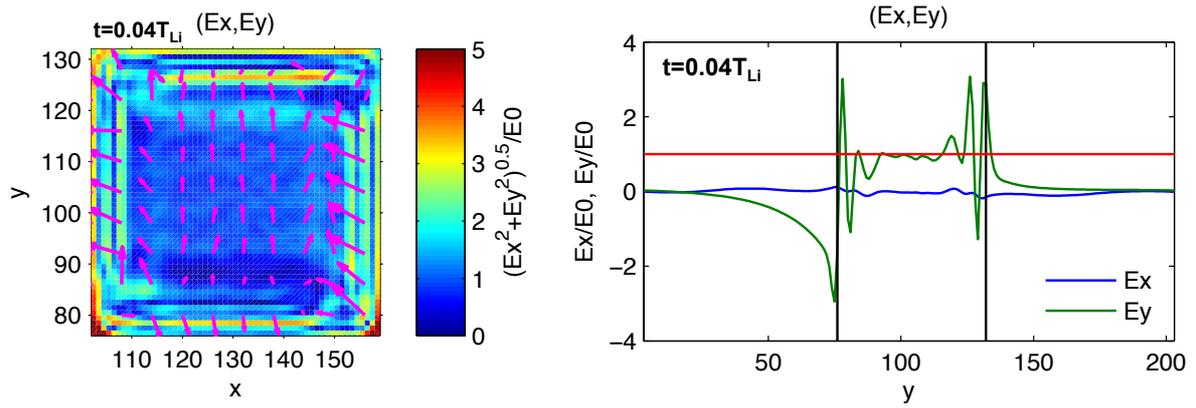

**Figure 7.** Case A: (Left) Intensity and orientation of the perpendicular electric field within the boundaries of the plasma element marked with a black rectangle in the left panels of Figure 6, at $t$=0.04$T_{Li}$; $T_{Li}$ is the ion Larmor period at the left hand side of the transition region. (Right) Variation profile of $E_x$ and $E_y$ along the $y$-axis, at $t$=0.04$T_{Li}$. The two black vertical lines indicate the boundaries of the plasma element along the $y$-axis.

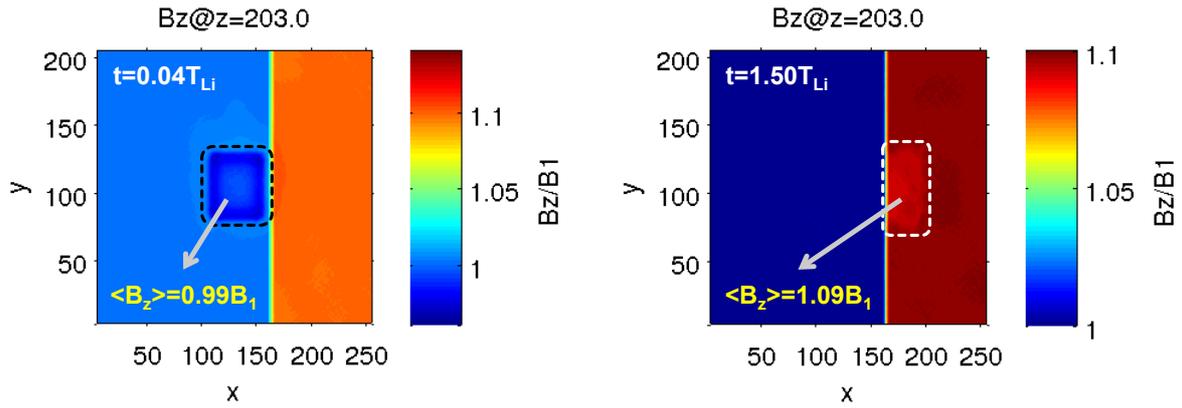

**Figure 8.** Case A: $B_z$ component of the total magnetic field in the $z$=203 cross-section inside the simulation domain, at $t$=0.04$T_{Li}$ (left panel) and $t$=1.50$T_{Li}$ (right panel); $T_{Li}$ is the ion Larmor period at the left hand side of the transition region. The dashed rectangles illustrate the current boundaries of the plasma element.

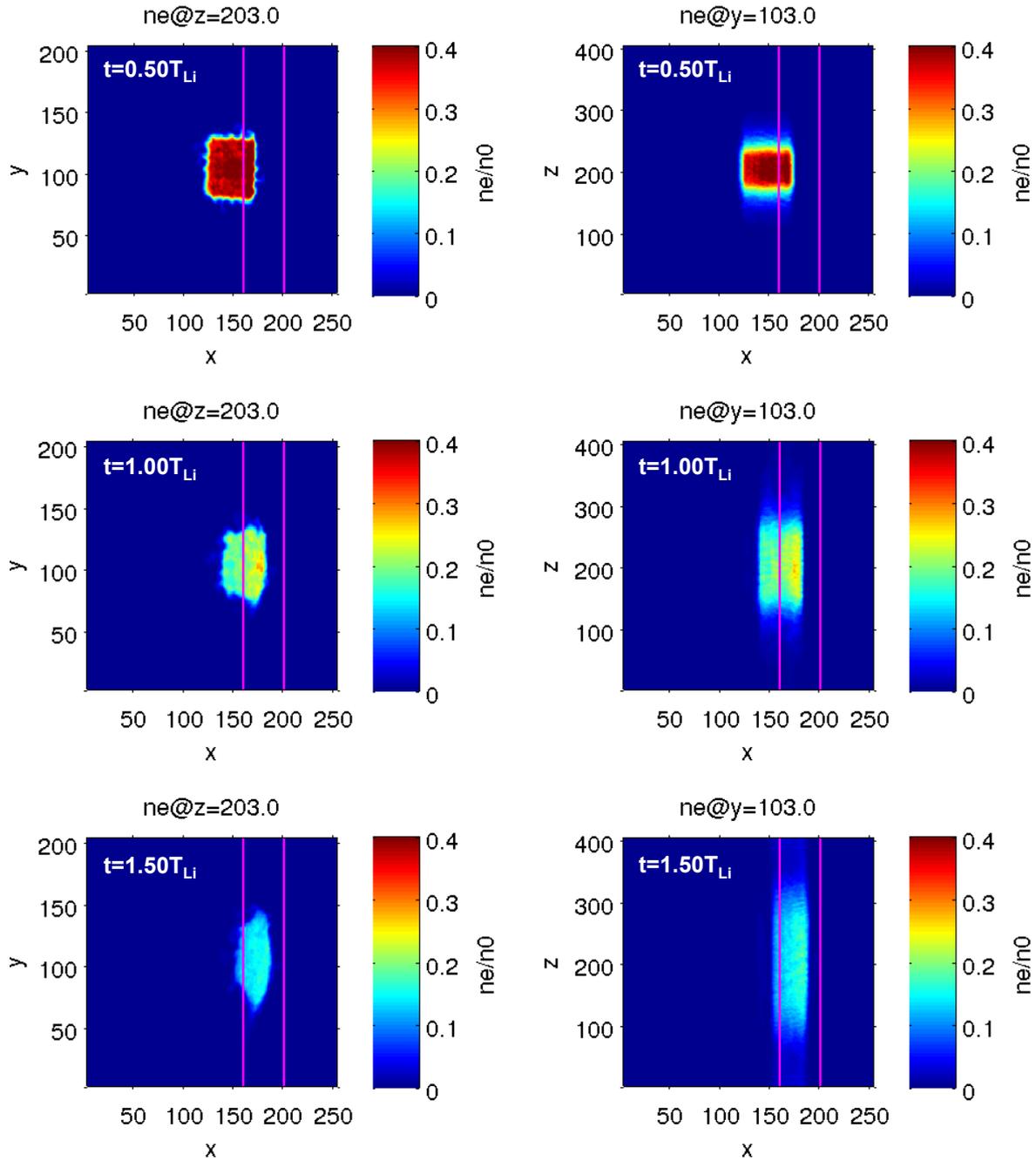

**Figure 9.** Case B: Number density of electrons in the $z$=203 (left column) and $y$=103 (right column) planes, at $t$=0.50$T_{Li}$ (top panels), $t$=1.00$T_{Li}$ (middle panels) and $t$=1.50$T_{Li}$ (bottom panels); $T_{Li}$ is the ion Larmor period at the left hand side of the transition region. The two magenta lines mark the boundaries of the transition region.

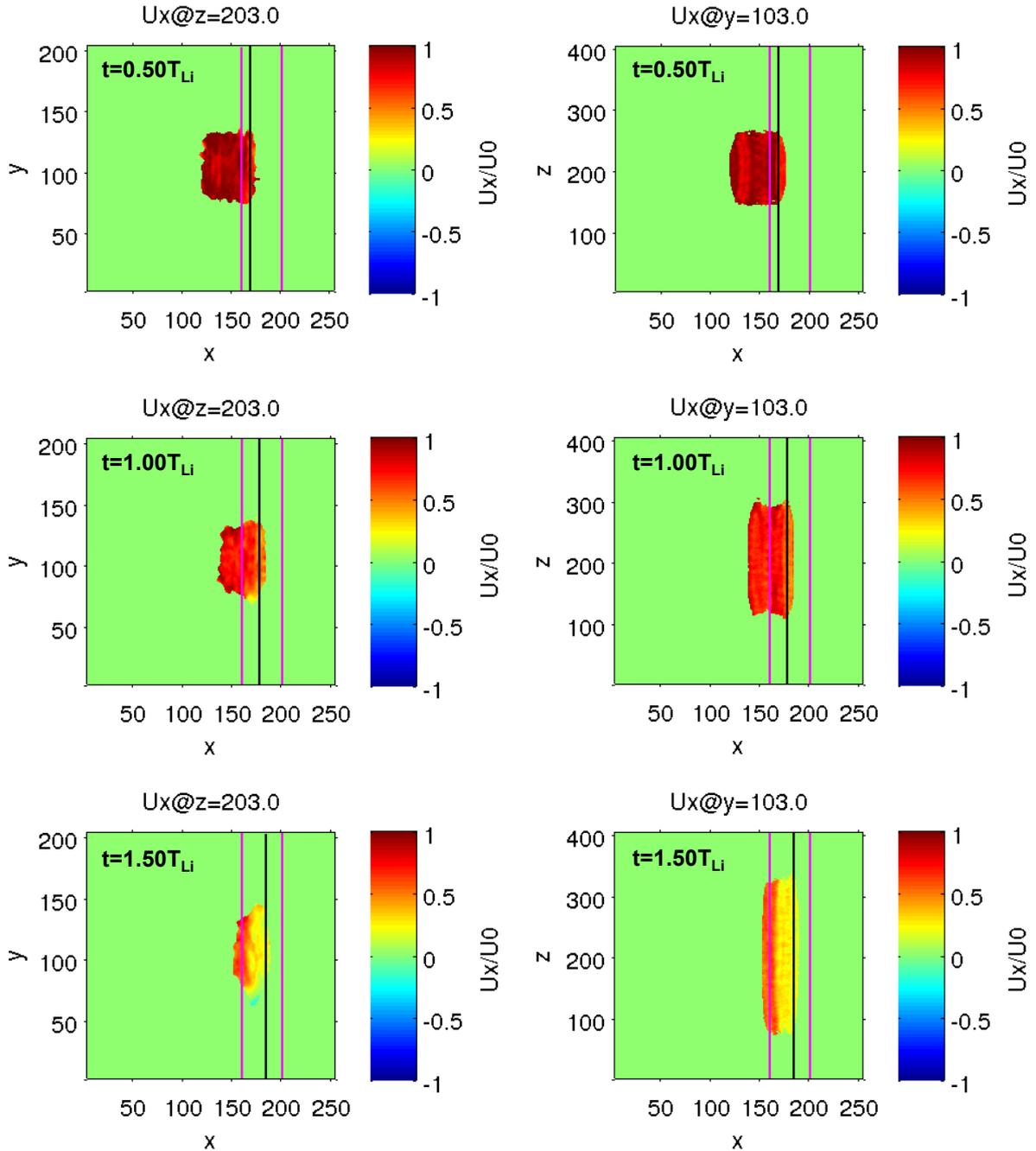

**Figure 10.** Case B: Normal component of the plasma bulk velocity, $U_x$, in the $z$=203 (left column) and $y$=103 (right column) planes, at $t$=0.50$T_{Li}$ (top panels), $t$=1.00$T_{Li}$ (middle panels) and $t$=1.50$T_{Li}$ (bottom panels); $T_{Li}$ is the ion Larmor period computed at the left hand side of the transition region. The two magenta lines mark the boundaries of the transition region.

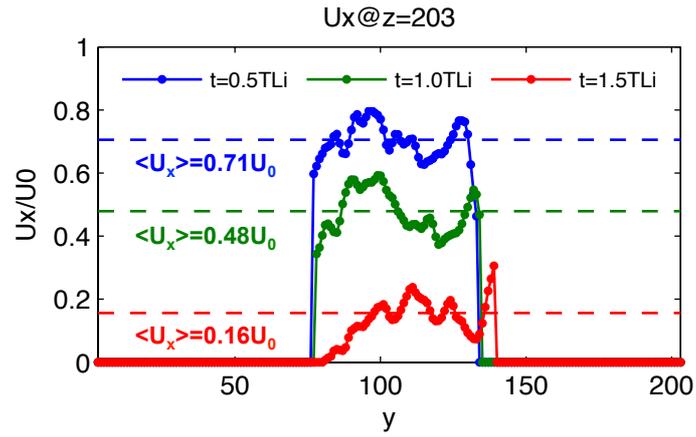

**Figure 11.** Case B: Variation profile of the normal component of the plasma bulk velocity, $U_x(y)$, at $t$=0.50$T_{Li}$ (blue line), $t$=1.00$T_{Li}$ (green line) and $t$=1.50$T_{Li}$ (red line), for the three black lines ($x$=168, $x$=177, $x$=183) shown in Figure 10; $T_{Li}$ is the ion Larmor period at the left hand side of the transition region. The dashed lines indicate the mean values of each profile shown.

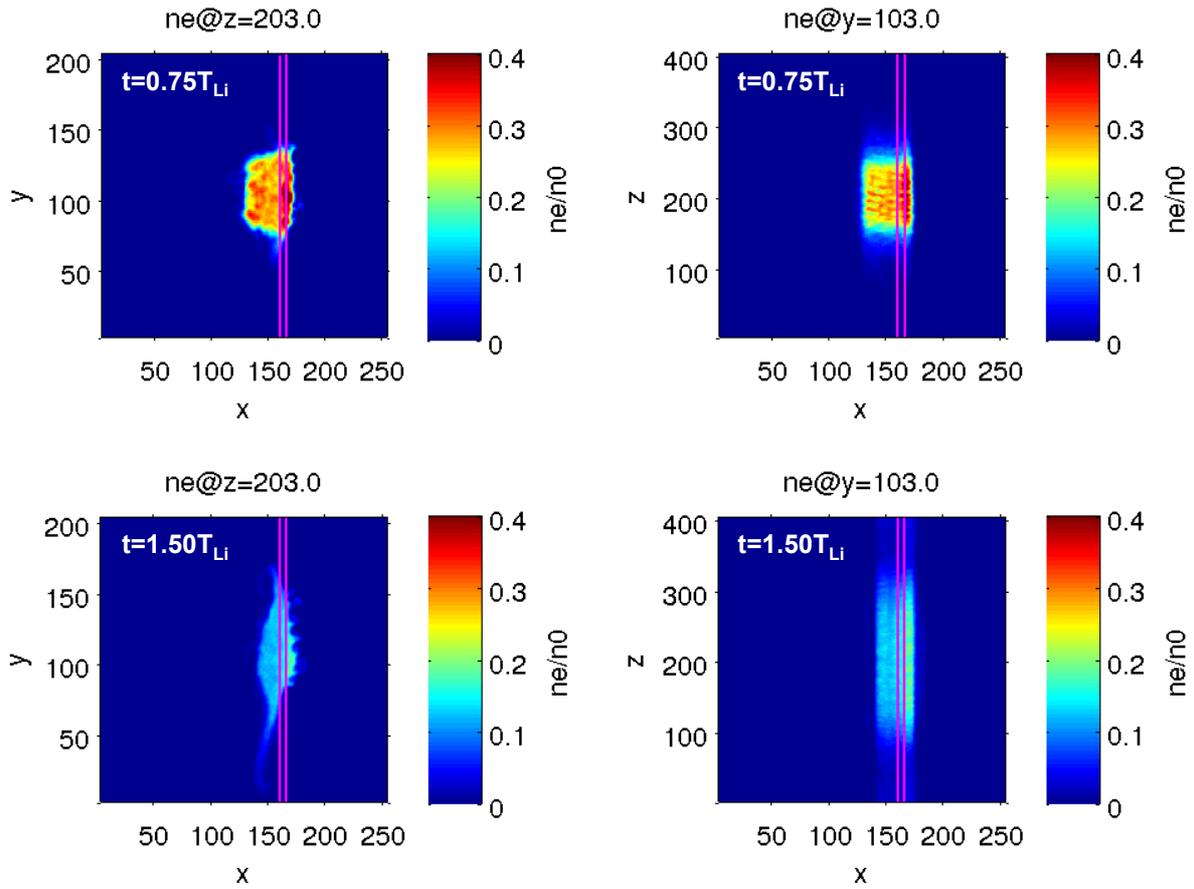

**Figure 12.** Case C: Number density of electrons in the *z*=203 (left column) and *y*=103 (right column) planes, at *t*=0.75$T_{Li}$ (top panels) and *t*=1.50$T_{Li}$ (bottom panels); $T_{Li}$ is the ion Larmor period at the left hand side of the transition region. The two magenta lines mark the boundaries of the transition region.

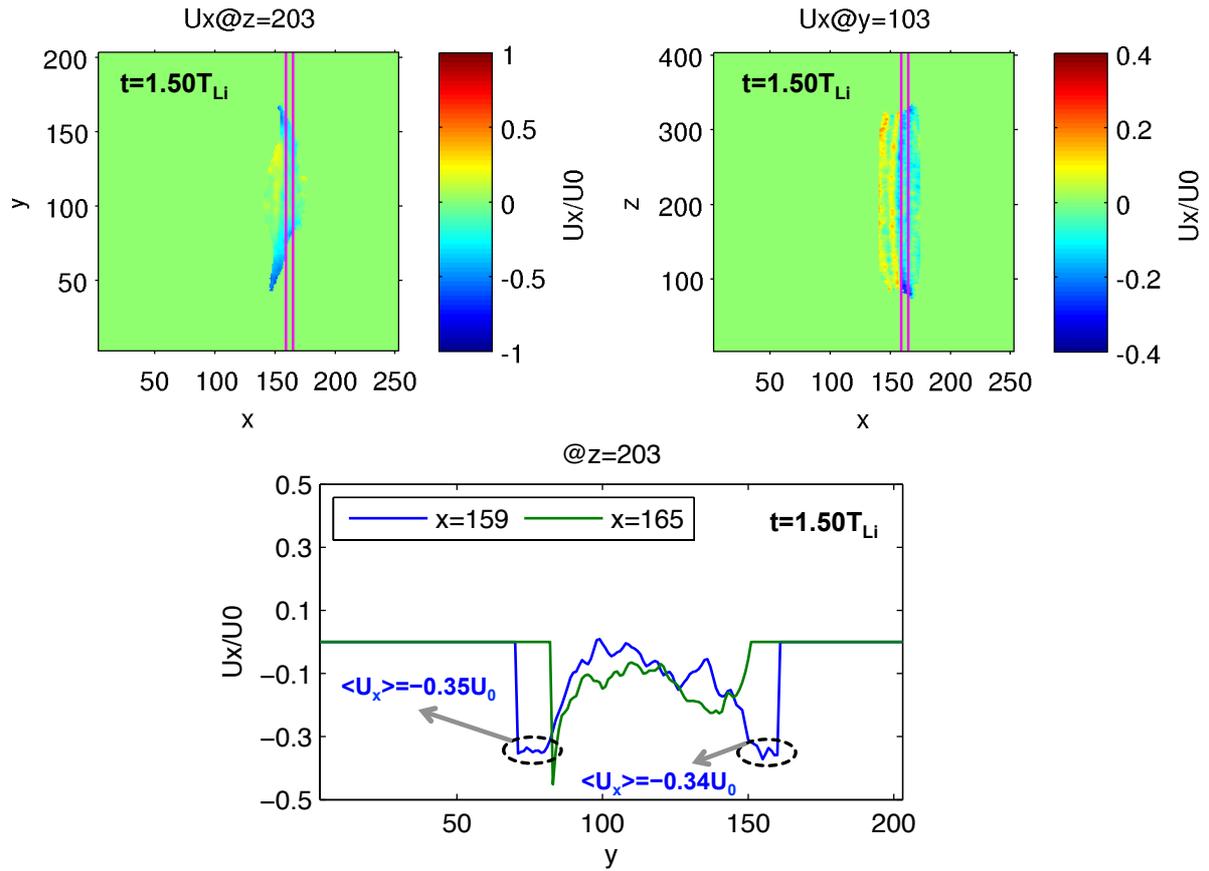

**Figure 13.** Case C: Normal component, $U_x$, of the plasma bulk velocity in the $z$=203 (top-left panel) and $y$=103 (top-right panel) planes, at $t$=1.50$T_{Li}$; $T_{Li}$ is the ion Larmor period at the left hand side of the transition region. The two magenta lines mark the boundaries of the transition region. The bottom panel shows the variation profile of $U_x$ along the $y$-axis, in $z$=203, for $x$=159 (blue line) and $x$=165 (green line), at $t$=1.50$T_{Li}$. The two dashed ovals illustrate the lateral "wings" of the plasma cloud.

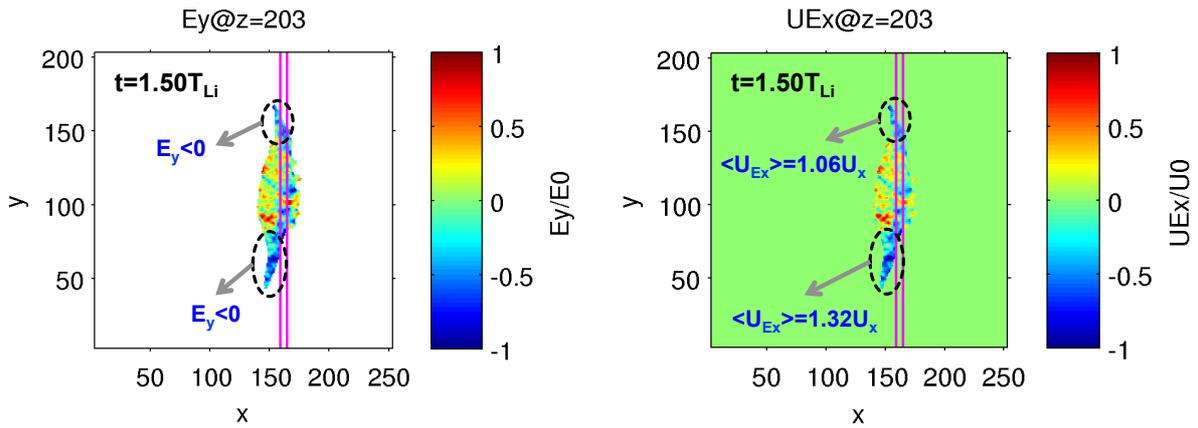

**Figure 14.** Case C: $E_y$ component of the electric field (left panel) and $U_{E,x}$ component of the electric drift velocity (right panel) in the $z$=203 cross-section inside the simulation domain, at $t$=1.50$T_{Li}$; $T_{Li}$ is the ion Larmor period at the left hand side of the transition region. The two magenta lines mark the boundaries of the transition region, while the dashed ovals illustrate the lateral "wings" of the plasma cloud.